\documentclass[a4paper]{article}

\usepackage{amsmath}
\usepackage{graphicx}
\usepackage{authblk}

\title{Intrinsic linewidth of quantum cascade laser frequency combs}

\author[1,2,$\flat$]{Francesco Cappelli}
\author[1]{Gustavo Villares}
\author[1]{Sabine Riedi}
\author[1,$\sharp$]{J\'er\^ome Faist}

\affil[1]{Institute for Quantum Electronics, ETH Zurich, 8093 Z\"urich, Switzerland}
\affil[2]{CNR-INO -- Istituto Nazionale di Ottica, Largo Enrico Fermi 6, 50125 Firenze FI, Italy \& \newline LENS -- European Laboratory for Non-Linear Spectroscopy, Via Nello Carrara 1, 50019 Sesto Fiorentino FI, Italy}

\affil[$\flat$]{francesco.cappelli@phys.ethz.ch}
\affil[$\sharp$]{jfaist@phys.ethz.ch}

\begin{document}

\maketitle

\begin{abstract}
The frequency noise power spectral density of a free-running quantum cascade laser frequency comb is investigated. A plateau is observed at high frequencies, attributed to the quantum noise limit set by the Schawlow-Townes formula for the total laser power on all comb lines. In our experiment, a linewidth of 292~Hz is measured for a total power of 25~mW. This result proves that the four-wave mixing process, responsible for the comb operation, effectively correlates the quantum noise of the individual comb lines.
\end{abstract}

\section*{Introduction}
\label{sec:intro}

In recent years, optical frequency combs (OFCs) have become fundamental tools for near infrared (NIR) spectroscopy and metrology \cite{Udem:1999a,Diddams:2000}. Thanks to their wide spectral coverage, high coherence and absolute traceability, they are used for high resolution and precision atomic and molecular spectroscopy in this spectral region \cite{Maddaloni:2009}. Since the fundamental ro-vibrational transitions of simple molecules fall in the mid infrared (MIR), it is of particular interest to have OFCs operating in this spectral region. Until now, a well-established approach to satisfy this need consisted of directly transferring NIR OFCs emission to the MIR region through nonlinear processes, e.g. difference frequency generation using intense fiber-based NIR OFCs \cite{Ruehl:2012,Zhu:2013,Galli:2013d} or optical parametric oscillators \cite{Adler:2009,Vodopyanov:2011}. This approach guarantees good  spectral coverage and coherence, but requires complex and delicate experimental set-ups. Comb generation has also been achieved by parametric oscillation in high-Q micro resonators \cite{Kippenberg:2011}, with a spectral coverage recently extended to the MIR region \cite{Wang:2013,Griffith:2015}.

Quantum cascade lasers (QCLs) \cite{Faist:1994} are current-driven semiconductor lasers based on intersubband transitions in quantum wells, emitting MIR or terahertz radiation. In devices designed with low group velocity dispersion, it has been shown that comb operation can be achieved \cite{Hugi:2012,Burghoff:2014} thanks to the four-wave mixing process taking place in the gain medium itself \cite{Friedli:2013}. For MIR-operating devices, the upper state lifetime, inherent to the intersubband transition of the active region, is very short (sub-picosecond range). This is responsible both for the broadband nature of the four-wave mixing process that enhances the mode-locking, but also for a tendency to 
operate with a mostly constant output power. For these reasons, the phase relation between the modes is similar to that of frequency-modulated lasers \cite{Hugi:2012}, as predicted by theory \cite{Khurgin:2014,Villares:2015}, and no pulses are emitted.

Quantum cascade laser frequency combs (QCL-combs) were initially characterized by measuring the autocorrellation of the intermode beat note at the cavity round-trip frequency (7.5~GHz), performing a so-called beat note spectroscopy~\cite{Hugi:2012}. A more sensitive technique is provided by comparing two QCL-combs in a heterodyne beat experiment. Recent experiments on QCLs in a dual-comb spectroscopy set-up demonstrated a mode equidistance as good as $7.5 \times 10^{-16}$ relative to the carrier optical frequency~\cite{Villares:2015}, a value close to those measured for micro-resonator-based combs~\cite{Kippenberg:2011}.

In this manuscript we investigate the frequency noise of such combs. This characterization is essential both for spectroscopy applications as well as for a better understanding of the fundamental properties of these devices. The generation of the comb of frequencies is interpreted within the framework of supermodes. A high-finesse optical cavity has been used as a multimode frequency-to-amplitude converter to retrieve the QCL-comb intrinsic linewidth. A comparison between the linewidth obtained in the comb regime operation and in the single-mode operation is also given, demonstrating that the four-wave mixing process effectively correlates the quantum noise of the comb lines.

\section{Frequency noise}
\label{sec:freqNoise}

What distinguishes a frequency comb from a simple array of perfectly equally spaced single-frequency optical sources is the correlation of the frequency noise. While the heterodyne beat of two independent single-frequency laser sources always yields a linewidth wider than that of the individual lasers, this is not the case if the two single frequencies are extracted from a frequency comb source. These considerations are equally true for technical as well as for quantum noise. The intrinsic linewidth of a laser is given by the Schawlow-Townes formula \cite{Schawlow:1958} and can be interpreted as the ratio of the number of photons emitted in the cavity by spontaneous emission over the number of the ones emitted by stimulated emission. As compared to a single-frequency device, the only effect of comb operation is the redistribution of the stimulated photons into equally spaced modes. For this reason, the intrinsic linewdith of a single comb line is expected to be unchanged and can be expressed by the Schawlow-Townes formula considering the total optical power of all comb lines.

As in micro-resonators-based combs, QCL-combs are generated through a four-wave mixing process. For this reason, starting from a semiclassical approach for QCL-combs \cite{Khurgin:2014,Villares:2015}, it makes sense to compare the quantities with the quantum formalism developed for micro-resonators-based combs \cite{Chembo:2014} to retrieve the Langevin equation for the photon annihilation operator related to the $n$-th QCL-comb line:
\begin{equation}
\begin{aligned}
\dot{\hat{a}}_n =& \left(\frac{G_n - 1}{2 \tau_c} + \textrm{i} D_n\right) \hat{a}_n \\
& - \frac{G_n}{2 \tau_c}  \sum_{k,l} C_{kl} B_{kl}~\hat{a}^\dagger_k \hat{a}_m \hat{a}_l~\kappa_{n,k,l,m} \\
& +  \frac{1}{\tau_c} \hat{V}_{n}
\label{eq:eq_operator_comb}
\end{aligned}
\end{equation}
where
\begin{align*}
G_n = & \frac{\textrm{i} \gamma_{12}}{2\pi~n~f_{\textrm{rep}} + \textrm{i} \gamma_{12}}~g_0 \\
D_n = & \frac{-\pi \delta_n^2}{f_n} - 2\pi~\delta_n \\
C_{kl} = & \frac{\gamma_{22}}{\gamma_{22} + 2\pi~\textrm{i}~(l-k)~f_{\textrm{rep}}} \\
B_{kl} = & \frac{\gamma_{12}}{2 \textrm{i}} \left(\frac{1}{-\textrm{i}\gamma_{12} - 2\pi~l~f_{\textrm{rep}}} - \frac{1}{\textrm{i}\gamma_{12} - 2\pi~k~f_{\textrm{rep}}} \right)
\label{eq:def_operator_comb}
\end{align*}
with $\kappa_{n,k,l,m}$ spatial superposition integral among the modes involved in the four-wave mixing process, $\tau_c$ photon lifetime in the laser cavity, $\gamma_{22}$ scattering rate out of the excited laser state, $\gamma_{12}$ loss of coherence of the laser transition, $f_{\textrm{rep}}$ comb line spacing (cavity round trip, without dispersion), $g_0$ peak gain, $\delta_n$ difference between the frequency of the $n$-th mode of the ideal laser cavity $f_n$ and the frequency of the $n$-th mode of the laser cavity with dispersion. $\hat{V}_n$ are the vacuum Langevin noise operators related to the optical loss processes (waveguide and mirrors) \cite{Benkert:1990,Chembo:2014}, characterized by the following statistical properties:
\begin{equation}
\begin{aligned}
& \langle \hat{V}_n (t) \rangle = 0 \\
& \langle \hat{V}_n^\dagger (t) \hat{V}_n (t') \rangle = \delta(t-t') \\
& \langle \hat{V}_n (t) \hat{V}_n (t') \rangle = 0 \\
\label{eq:noise_op_st_prop}
\end{aligned}
\end{equation}
The second term in equation~\ref{eq:eq_operator_comb} is due to the four-wave mixing and it is responsible for the coupling among all the laser modes. Solving equation~\ref{eq:eq_operator_comb} is beyond the scope of this paper. However, we note that in comb operation the average relative phases of the modes are fixed (within fluctuations). Therefore, through a unitary transformation, it is possible to select a new basis for the cavity modes, the \emph{supermodes} basis, such that one of this supermodes corresponds to that selected by the comb operation \cite{Haken:1968}. The new annihilation operators are given by
\begin{equation}
\hat{b}_q = \sum_n U^n_q \hat{a}_n
\label{eq:superm_annih}
\end{equation}
where $U^n_q$ is the element of a unitary matrix such that $\mathrm{U}^{-1} = \mathrm{U}^\dagger$ \cite{Grynberg:2010}. In this way, the equation is  reduced to that of a single-mode laser (with only one mode excited). In particular, the Langevin operators for the new modes
\begin{equation}
\hat{V}'_q = \sum_n U^n_q \hat{V}_n
\label{eq:noise_op_supermod}
\end{equation}
will have the same correlation properties (equations \ref{eq:noise_op_st_prop}) because of the unitary nature of the transformation. The resulting frequency noise is expected to be the same of a single-mode laser.

\section{Experiment}
\label{sec:experiment}

\begin{figure}[htbp]
\centering
\fbox{\includegraphics[width=0.5\linewidth]{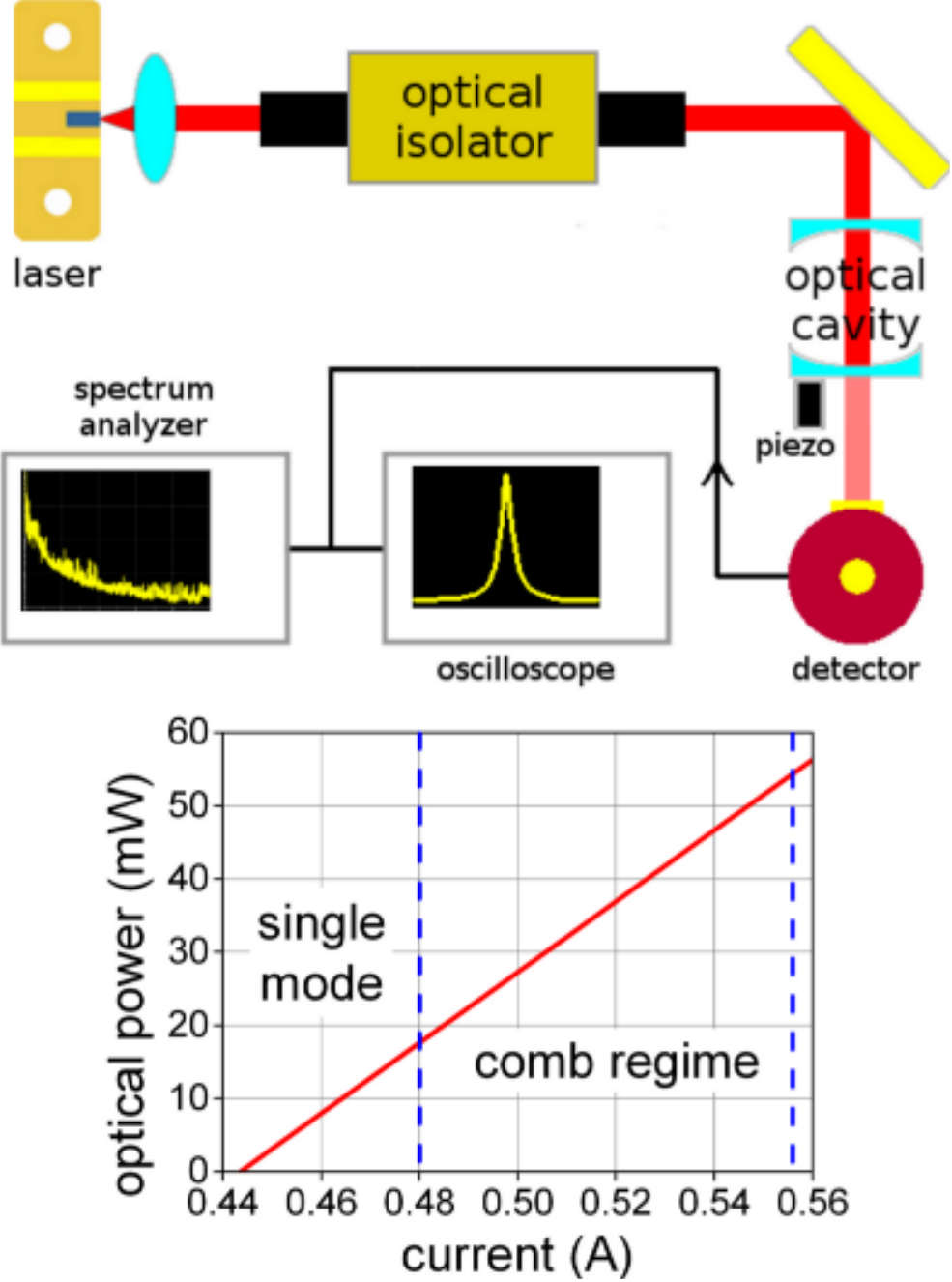}}
\caption{\textbf{Top:} Experimental set-up used to measure the FNPSD of the laser. The main optical components include the laser (a multistack InGaAs/InAlAs QCL), the optical isolator, the high-finesse optical cavity and the high-sensitivity MCT detector. The signal is processed by a high-sampling-rate oscilloscope. \newline
\textbf{Bottom:} Power-versus-current curve of the QCL at fixed temperature. Two operation regimes are observed in this device, a single-mode regime and a comb regime. }
\label{fig:cavSetupPow}
\end{figure}

\begin{figure}[htbp]
\centering
\fbox{\includegraphics[width=0.5\linewidth]{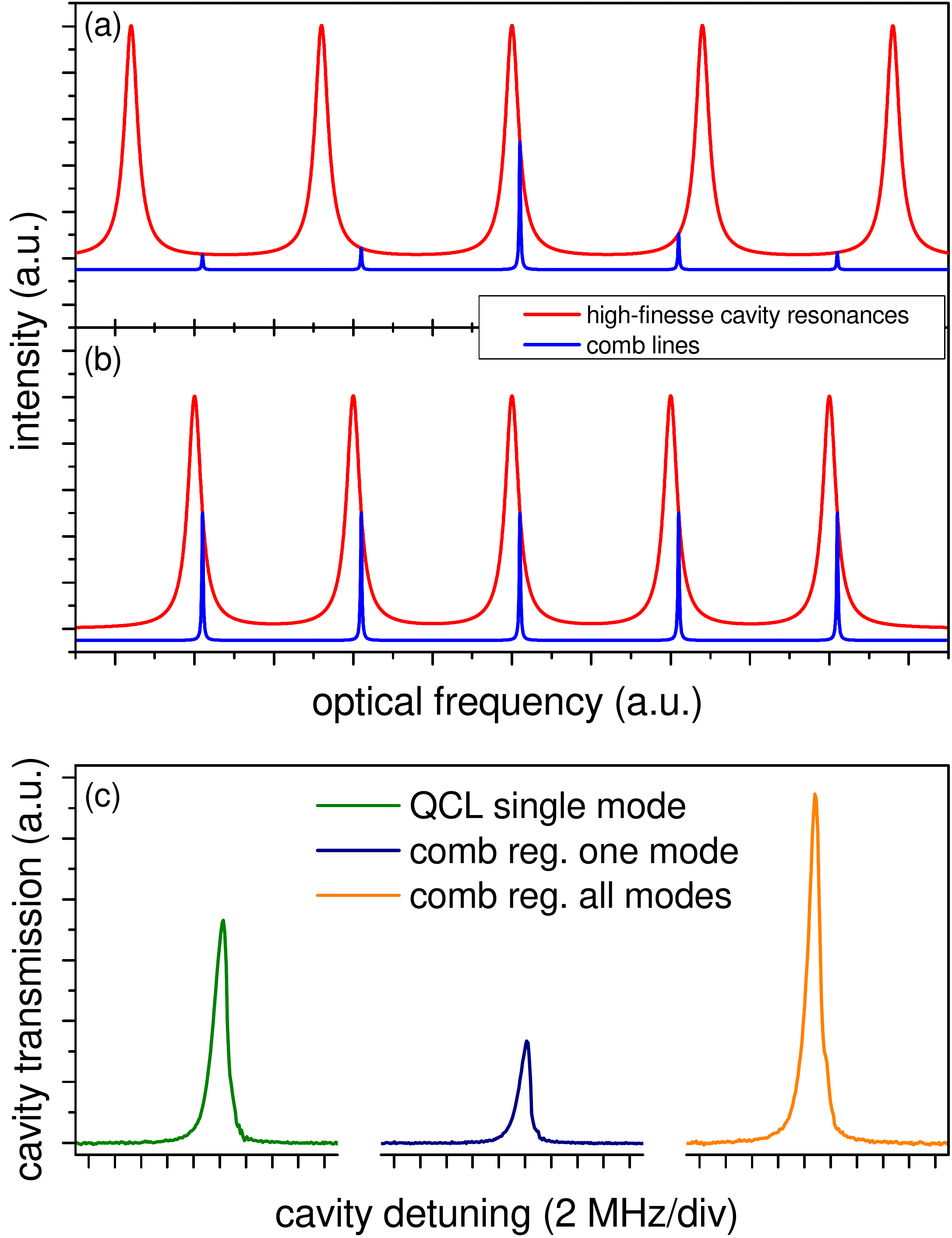}}
\caption{Schematic of the optical cavity and of the comb spectra with $Vr \neq 1.0$ (comb regime - one mode) and $Vr=1.0$ (comb regime - all modes) (\textbf{a)} and \textbf{b)} respectively). \newline
\textbf{c)} Cavity transmissions acquired in the three conditions: single-mode QCL, QCL in comb regime with only one mode in resonance with the cavity, and QCL in comb regime with all the modes in resonance with the cavity. These acquisitions are obtained by scanning the cavity length. The cavity detuning is the variation of the resonance frequency ($FSR$) with the length. These peaks are used for the calibration of the FA converter.
}
\label{fig:trasmCav}
\end{figure}

The laser used for these experiments is a QCL-comb based on an InGaAs/InAlAs broadband design with multiple active regions (multistack), previously reported in~\cite{Hugi:2012}. It operates in continuous wave at room temperature emitting several milliwatts of power at $7.10~\mu$m. The device length is $6$~mm, corresponding to $f_{\textrm{rep}} \approx 7.5$~GHz. Two main operation regimes are observed in this device. Just above threshold, the device emits single-mode radiation. A \emph{comb regime} is then observed for a significant part of the device working range (see figure \ref{fig:cavSetupPow}). This very convenient property enabled us to study and compare the linewidth in both regimes (single-mode and comb) in the same device.
In order to investigate the frequency noise power spectral density (FNPSD), a high-finesse optical cavity (\emph{Fabry-P\'erot}, $\mathcal{F} \approx 6000$) was used to resolve the laser spectrum and to detect the frequency fluctuations of the laser, acting as frequency-to-amplitude (FA) converter (figure~\ref{fig:cavSetupPow}). In the set-up, an optical isolator (transmission $T=-2.9$~dB, extinction $E=-33$~dB) has been used to avoid the instabilities induced on the laser by the back-reflection from the input mirror of the cavity. To collect the signal transmitted by the cavity, a high-sensitivity nitrogen-cooled HgCdTe (MCT) detector ($BW=0-10$~MHz) was used. A 12-bit vertical resolution, 1~GHz analog bandwidth, 2.5~GS/s sampling rate oscilloscope was used to collect the signal and to compute its Fourier transform. The distance between the two mirrors was chosen in order to set the free spectral range ($FSR$) of the cavity close to the comb line spacing $f_{\textrm{rep}}$. In order to resolve the laser spectrum, a \emph{Vernier ratio} $Vr=FSR/f_{\textrm{rep}}$ slightly different from one was chosen and a piezoelectric actuator was used to scan the cavity length over one $FSR$~\cite{Galli:2014a}. A schematic representation is depicted in figure~\ref{fig:trasmCav}a.

To utilize the cavity as a FA converter, we acted again on the piezoelectric actuator and on the temperature controller of the laser to set $Vr=1.0$ and to let the comb offset frequency $f_{\textrm{o}}$ be equal to the one of the optical cavity. In this way, the comb lines and the optical cavity resonances were perfectly matched (see figure~\ref{fig:trasmCav}b). As a consequence, in these conditions and \emph{only} in these conditions of temperature and driving current of the laser, all the comb lines were transmitted by the cavity. The cavity can thus be used as a \emph{multimode} frequency-to-amplitude converter to collect the frequency fluctuations of all the modes at the same time \cite{Galli:2013d} (see Supplement \ref{sec:multi_freqAmp_conv} for a demonstration). Since $Vr=1.0$, an accurate value of the optical cavity free spectral range $FSR$ can be obtained by measuring $f_{\textrm{rep}}$. The latter was measured by acquiring the intermode beat note on the laser current with a spectrum analyzer \cite{Villares:2014}. An accurate value of the cavity $FSR$ is needed for the calibration of the FA converter. Under optimal conditions of current and temperature for the matching, the laser emits a power $P=25$~mW. Thanks to the high finesse, it is also possible to collect the frequency fluctuations of an individual comb line by slightly varying the $FSR$. A spectrum retrieved with the laser in single-mode operating conditions ($P=15$~mW) was also acquired.
\begin{figure}[htbp]
\centering
\fbox{\includegraphics[width=\linewidth]{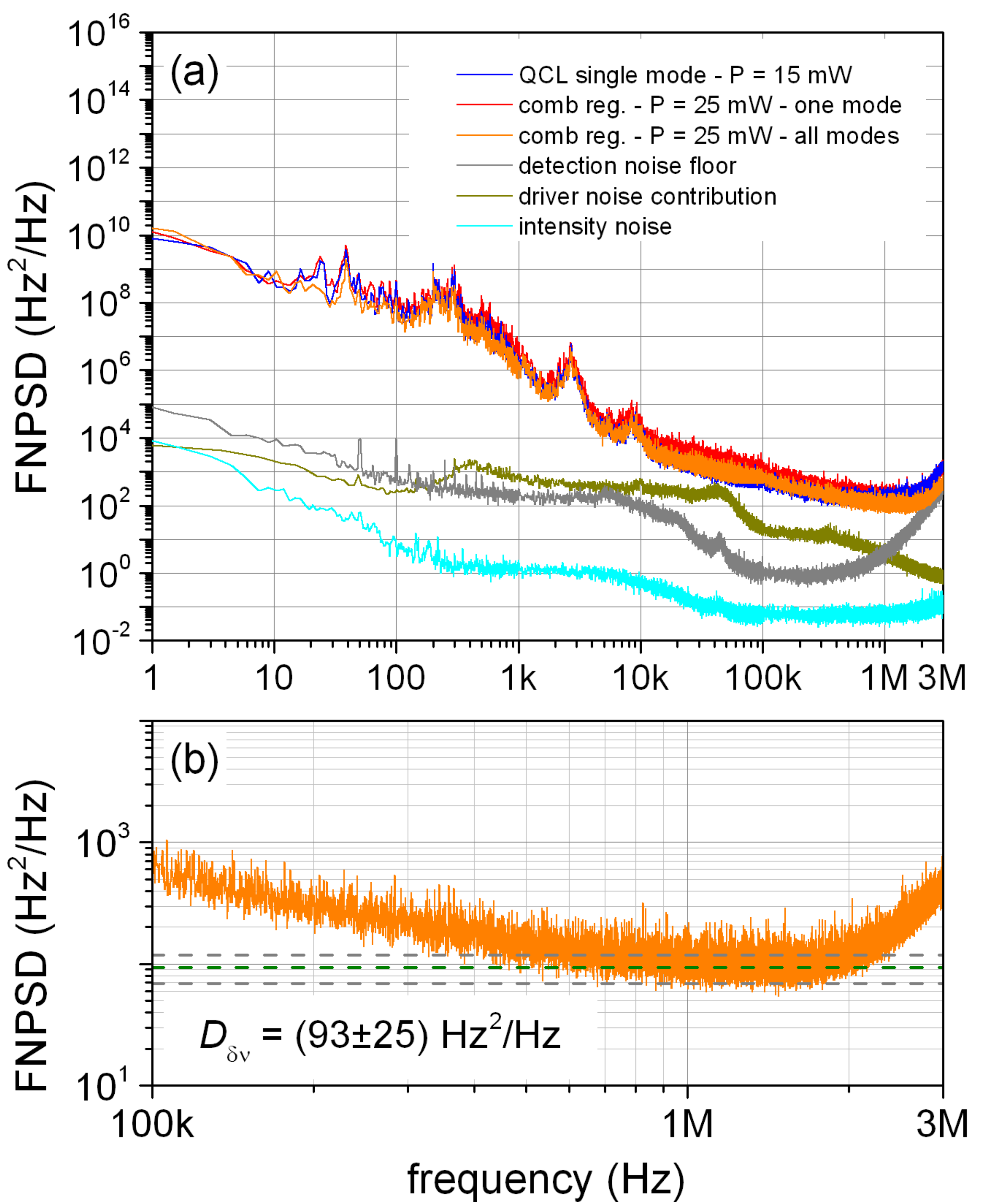}}
\caption{\textbf{a)} FNPSD of the QCL-comb taken in three different conditions. The spectra are compensated for the FA converter cutoff. The technical contributions to the noise are also reported: taking into account the detection noise floor shape, the spectra are reliable up to 3~MHz; the two contributions given one by the current driver and the other one related to the intensity noise are negligible. \newline
\textbf{b)} Zoom of the flattening portion of the spectrum around 1~MHz, corresponding to the Schawlow-Townes limit. }
\label{fig:FNPSD}
\end{figure}
The FNPSD measured on the single-mode regime as well as on the comb regime are reported in figure~\ref{fig:FNPSD}a after correction for the finite bandwidth of the FA converter (response frequency cutoff of the high-finesse cavity related to the photon lifetime). We observe that the frequency noise on the comb regime is identical to the frequency noise on the single-mode regime. Moreover, the frequency noise of an individual comb line is also equivalent to the one acquired on all the comb lines together. By integrating the FNPSD using Elliott's formula \cite{Elliott:1982}, the full width at high maximum (FWHM) of a laser mode can be retrieved. In our case, we obtain a FWHM of about 600~kHz in a 1-s timescale, which is consistent with the linewidth shown by distributed-feedback (DFB) QCLs \cite{Bartalini:2011,Tombez:2011,Cappelli:2012}. Moreover, the contributions of the driver noise and the laser intensity noise as well as the detection noise floor are reported. Taking into account the detection noise floor shape, the spectra are reliable up to 3~MHz. Around 1~MHz, a flattening can be observed. Figure~\ref{fig:FNPSD}b shows a portion of the same FNPSD spectrum (from 100~kHz to 3~MHz). This flattening, characteristic of a white frequency noise, corresponds to the intrinsic quantum noise level $D_{\delta \nu}$ due to the spontaneous emission, the so called \emph{Schawlow-Townes limit} \cite{Schawlow:1958}.

\section{Discussion}
\label{sec:discussion}

It is interesting to compare this level $D_{\delta \nu}$ to the one expected for a single-mode device with the same characteristics \cite{Henry:1982}:
\begin{equation}
\delta \nu = \frac{h \nu}{P} \frac{\alpha_{\textrm{tot}} c^2}{4 \pi n_{\textrm{g}}^2} \alpha_{\textrm{m}} n_{\textrm{sp}} (1+\alpha_{\textrm{e}}^2)
\label{eq:SchToDnA}
\end{equation}
Taking $\nu=42.2$~THz as central frequency, $P=25$~mW as total power emitted by the laser, $\alpha_{\textrm{m}}=2.2~\textrm{cm}^{-1}$ as mirror losses, $\alpha_{\textrm{tot}}=7.2~\textrm{cm}^{-1}$ as total losses (the relatively high waveguide losses are due to the residual \emph{cross-absorption} given by the multistack structure), $n_{\textrm{g}}=3.4$, $n_{\textrm{sp}}=2$ as spontaneous emission factor and $\langle \alpha_{\textrm{e}}^2 \rangle = 0.0023$ as squared \emph{Henry linewidth enhancement factor} averaged over the laser spectrum (see Supplement \ref{sec:Henry_est}), we obtain $\delta \nu = 230$~Hz, value which is consistent with the one obtained from the spectrum $\delta \nu = \pi D_{\delta \nu} = (292 \pm 78)$~Hz (figure~\ref{fig:FNPSD}b). This justifies the theoretical framework introduced in section \ref{sec:freqNoise}.

More importantly, the measurement of the FNPSD in comb regime shows that the quantum fluctuations of the different modes are correlated. In fact, we observe that the FNPSD -- in particular the portion limited by the quantum noise -- is identical when measured with one comb line and with all comb lines simultaneously. This quantum limit -- value which is given by the Schawlow-Townes expression -- would be at least a factor of 6 larger than the one showed in figure \ref{fig:FNPSD}b, assuming that the quantum fluctuations of each comb mode are uncorrelated. This factor is outside our uncertainty.

\section*{Conclusions}
\label{sec:conclusions}

With this work we have demonstrated that in quantum cascade laser frequency combs the four-wave mixing process at the origin of the comb operation regime is also responsible for the correlation between the frequency fluctuations between the modes until the quantum limit. As a result, the linewidth is shown to be limited by the Schawlow-Townes formula, as it is for single-mode lasers of the same total power. As a consequence, instruments using the spectral multiplexing of dual-combs or multi-heterodyne spectrometers hold an inherent noise advantage compared to similar systems using arrays of single-mode lasers. Finally, the same technique used to retrieve the frequency noise power spectral density could be used to implement an active stabilization for locking this combs to high-finesse ultra-stable optical cavities.

\section*{Funding Information}
This work was financially supported by the Swiss National Science Foundation, the ETH Pioneer Fellowship programme, the European Laboratory for non-linear Spectroscopy (Florence) and the Italian National Institute of Optics (CNR-INO).

\section*{Acknowledgments}

We thank P. De Natale for useful discussions.



\newpage

\section{Intrinsic linewidth of quantum cascade laser frequency combs: supplementary material}
\label{sec:supplementary}

\subsection{Demonstration of the multimode frequency-to-amplitude conversion}
\label{sec:multi_freqAmp_conv}

In this section we will show how satisfying the right conditions it is possible to use an optical cavity as \emph{multimode} frequency-to-amplitude converter to retrieve the frequency noise of an optical frequency comb. The intensity of the $i$-th comb line transmitted by the optical cavity is given by
\begin{equation}
I_i = I_{i0} T(\nu_{ij})
\label{eq:transmSingTooth}
\end{equation}
where $I_{i0}$ is the intensity of the $i$-th comb line, $\nu_{ij} = \nu_j - \nu_i$, $\nu_i$ is the center frequency of the $i$-th comb line, $\nu_j$ is the center frequency of the $j$-th cavity transmission and $T(\nu_{ij})$ is the convolution between the $i$-th comb line shape and the $j$-th cavity transmission profile. The total intensity transmitted by the cavity is given by
\begin{equation}
I_\textrm{tot} = \sum_i I_i = \sum_i \left[ I_{i0} T(\nu_{ij}) \right]
\label{eq:transmTOTgen}
\end{equation}
If
\begin{equation}
\nu_{ij} = \Delta \nu = \textrm{const.} \qquad \forall i,j
\label{eq:transmCondi}
\end{equation}
we are allowed to bring $T(\nu)$ out of the sum, yielding
\begin{equation}
I_\textrm{tot} = T(\Delta \nu) \sum_i I_{i0} = T(\Delta \nu) I_{0~\textrm{tot}}
\label{eq:transmTOTeq}
\end{equation}
This proves that in this conditions the transmission and consequently the frequency-to-amplitude conversion follows the same rules of the single-mode case. Actually, the condition expressed by equation \ref{eq:transmCondi}, which corresponds to impose that both the comb lines and the cavity resonances have the same dispersion in frequency, 
is not so strict. It is in fact sufficient that
\begin{equation}
| \nu_{ij} - \nu_{lm} | \ll \frac{1}{2} \mathrm{FWHM} \left[ T(\Delta \nu) \right]
\label{eq:lessLin}
\end{equation}
for all the transmitted peaks. In other words it is sufficient that the accumulated dispersion over the whole comb spectrum is small compared to the width of the cavity resonances.

\subsection{Henry linewidth enhancement factor estimation}
\label{sec:Henry_est}

In order to compute the Schawlow-Townes limit
\begin{equation}
\delta \nu = \frac{h \nu}{P} \frac{\alpha_{\textrm{tot}} c^2}{4 \pi n_{\textrm{g}}^2} \alpha_{\textrm{m}} n_{\textrm{sp}} (1+\alpha_{\textrm{e}}^2)
\label{eq:SchToDnB}
\end{equation}
a proper estimation of the Henry linewidth enhancement factor is needed \cite{Henry:1982}. As reported in \cite{Faist:2013bo}, assuming a Lorentian lineshape for the intersubband transition, the Henry factor for a QCL can be expressed as
\begin{equation}
\alpha_{\textrm{e}}(\nu)=\frac{\nu_{32}-\nu}{\gamma_{32}}
\label{eq:HenFac}
\end{equation}
where $\nu_{32}$ and $\gamma_{32}$ are the center frequency and the half with at high maximum of the transition respectively, while $\nu$ is the frequency of the radiation emitted by the laser. Concerning $\nu$, we have to consider that the emitted radiation is a comb of frequencies, so $\alpha_{\textrm{e}}$ has to be computed as average over the whole emitted spectrum
\begin{equation}
\langle \alpha_{\textrm{e}}^2 \rangle_\textrm{laser spectrum} = \frac{1}{\Delta \nu~\gamma_{32}^2} \int_\textrm{las. sp.} (\nu_{32}-\nu')^2~\textrm{d} \nu'
\label{eq:avgHenFacInt}
\end{equation}
where $\Delta \nu$ is the width of the comb spectrum. Assuming that the center of the emitted spectrum corresponds to $\nu_{32}$ we obtain directly
\begin{equation}
\langle \alpha_{\textrm{e}}^2 \rangle_\textrm{laser spectrum} = \frac{1}{3 \gamma_{32}^2} \left(\frac{\Delta \nu}{2} \right)^2
\label{eq:avgHenFacFin}
\end{equation}
Finally, knowing from electroluninescence measurements that $\gamma_{32} = 150~\textrm{cm}^{-1}$ and that $\Delta \nu = 25~\textrm{cm}^{-1}$ in the actual operating conditions of the laser ($P = 25$~mW) we have
\begin{equation}
\langle \alpha_{\textrm{e}}^2 \rangle_\textrm{laser spectrum} = 0.0023
\label{eq:avgHenFacVal}
\end{equation}
value that can be neglected in equation \ref{eq:SchToDnB}. In case of a multistack device the shape of the gain curve on the top is even flatter than a Lorentian, consequently the given value for $\langle \alpha_{\textrm{e}}^2 \rangle$ has to be considered as an upper limit.

\end{document}